
\font\bigbold=cmbx12
\font\smc=cmcsc10
\magnification=\magstep1
\vsize=23.5 truecm
\hsize=16.8 truecm
\hoffset=-.5truecm
\baselineskip=12pt
\parskip=5pt
\nopagenumbers
\parindent=0pt
\null
\vskip 2truecm
\centerline{\bigbold Reply to HEPTH-9509028}
\medskip
\vskip 30pt
\centerline{\smc Martin Lavelle}
\vskip 5pt
{\baselineskip=13pt
\centerline{Grup de F\'isica Te\`orica and IFAE}
\centerline{Edificio Cn}
\centerline{Universitat Aut\'onoma de Barcelona}
\centerline{E-08193 Bellaterra (Barcelona)}
\centerline{Spain}
\centerline{(e-mail: lavelle@ifae.es)}
}
\vskip 10pt
\centerline
{\smc and}
\vskip 8pt
\centerline{\smc David McMullan}
\vskip 5pt
{\baselineskip=13pt
\centerline{School of Mathematics and Statistics}
\centerline{University of Plymouth}
\centerline{Drake Circus, Plymouth, Devon PL4 8AA}
\centerline{U.K.}
\centerline{(e-mail: d.mcmullan@plymouth.ac.uk)}
}

\vskip 3.5truecm
\noindent Much is known about the charged sectors of Quantum
Electrodynamics (QED). The state space displays rich superselection
sectors where, in particular, the velocity of charged particles, like
the electron, labels different sectors$^{[1,2]}$. The BRST charge can
be used effectively to project out the gauge invariant, physical states,
but the resulting state space is an infinite tensor product of these
superselection sectors, i.e., the BRST charge is not capable, on its
own, of distinguishing between these physically distinct sectors. In
our letter$^{[3]}$ we introduced a new, non-local, non-covariant
symmetry that projected out the static sector of QED (described by
Dirac's dressed electron). The non-static extension of Dirac's
construction are still BRST, and indeed anti-BRST, invariant, but
are {\it not} invariant under our symmetry. As such our symmetry,
in conjunction with BRST, can be used to extract an irreducible
sector of the intricate structure of QED. Indeed, as one would expect to
be able to recover the other, velocity dependent sectors in a similar
fashion, we conjecture that our symmetry is but one member of a whole
class of distinct, new, non-covariant symmetries of QED.
\goodbreak

In our presentation of the symmetry we took great pains to show that
it really is a symmetry of QED. In the comment a simplistic argument
is presented in {\it free} QED (which is, after all, a trivial theory
without interaction or charges and which hence lacks the above
superselection sectors) to claim that our symmetry is no more than
a \lq\lq  a non-local
version of standard BRST\rq\rq. No matter what the merits of these
arguments are for the free theory, this is manifestly not true for
the interacting theory.  The comment exemplifies the dangers of
formal manipulations of path integrals in non-trivial theories
like the full version of QED.

On a lesser note, we point out that Tang and Finkelstein's work$^{[4]}$
is not a covariant version of our symmetry; as far as we are aware this
has also never been suggested by those authors.

\bigskip\bigskip
\noindent {\bf References:}
\bigskip
\item{[1]}{R.\ Haag, {\sl Local Quantum Physics}, (Springer-Verlag,
Berlin, Heidelberg, 1993).}
\item{[2]}{D.\ Buchholz, {\sl Commun.\ Math.\ Phys.}\ {\bf 85}
(1992) 49.}
\item{[3]}{M.\ Lavelle and D.\ McMullan, {\sl Phys.\ Rev.\ Lett.}\
{\bf 71}
(1993) 3758.}
\item{[4]}{Z.\ Tang and D.\ Finkelstein, {\sl Phys.\ Rev.\ Lett.}\
{\bf 73} (1994) 3055.}

\bye